\documentclass{JHEP3}
\input epsf

\def\be{\begin{equation}}
\def\bea{\begin{eqnarray}}
\def\ee{\end{equation}}
\def\eea{\end{eqnarray}}

\def\la{\lambda}
\def\eps{\epsilon}

\preprint{
{\tt hep-th/0404006}}
\title{Adding momentum to D1-D5 system}
\author{Oleg Lunin\\
~~~~~~~~~~~~~~\\
School of Natural Sciences\\
Institute for Advanced Study\\
 Einstein Drive, Princeton NJ 08540  USA\\
~~~~~~~\\}

\abstract{
We construct the first example of asymptotically flat solution which 
carries three charges (D1,D5 and momentum) and which is completely 
regular everywhere. The construction utilizes the relation between 
gravity solutions and spectral flow in the dual CFT. We show that the 
solution has the right properties to describe one of the microscopic 
states which are responsible for the entropy of the black hole with 
three charges.  
}

\begin{document}
\section{Introduction.}

One of the most fascinating problems in theoretical physics is the problem
of black hole entropy. Ever since Bekenstein first proposed his famous 
formula \cite{bek} relating the entropy of black hole with the area of 
the horizon, there were countless attempts to identify the microscopic 
states responsible for such entropy. A big step towards understanding 
of microscopic states was made in \cite{strVafa}. For a specific type 
of black holes which arises in string theory and which has three charges 
(corresponding to D1 branes, D5 branes and momentum), Strominger and Vafa 
counted the 
microscopic states of the branes and demonstrated perfect agreement with 
Bekenstein formula. This counting was later extended to the 
rotating \cite{bmpv} and near extremal \cite{cmhs} black holes.

While the work of Strominger and Vafa \cite{strVafa} explained how to 
{\it count} the states of the black hole, it did not address the question 
how to {\it see} them as different states in supergravity. Unfortunately 
the gravity picture for the microscopic states is still missing.

It is interesting that for a somewhat simpler system which has only two 
charges (D1 and D5, but no momentum), one can actually construct the 
geometries corresponding to {\it all} microscopic states \cite{lmPar,lmm}.
This system does not form a black hole, but it still has many degenerate 
microscopic states ($\sim \exp(2\sqrt{2}\pi\sqrt{n_1n_5})$ of them). In 
\cite{lmPar} a large class of these microscopic states was considered and 
the geometries corresponding to each of those states were constructed
\footnote{For earlier work on geometries of the two charge system see 
for example \cite{add2ch}.}. 
In \cite{lmm} this procedure was repeated for the remaining 
states\footnote{The five dimensional black hole is constructed by 
compartifying string theory on either $T^4$ or K3. The work \cite{lmPar} 
dealt with microstates which are universal and appear for both 
compactifications. In \cite{lmm} the states associated with excitations 
on $T^4$ were constructed explicitly, and the procedure for constructing 
excitations associated with K3 was outlined.} , and more importantly, 
it was proven that geometries corresponding to all microscopic states 
are completely regular. As we mentioned, D1--D5 system does not have a 
horizon, but it does have degenerate states and thus it does have entropy.
The way to explain an entropy for systems without horizons was proposed 
by Sen \cite{stretch} and it was based on a concept of a stretched 
horizon. It is interesting that for D1--D5 system one can see the 
emergence of an effective stretched horizon from the geometries which 
are completely regular, and the area of this stretched horizon reproduces 
the correct expression for the entropy \cite{lmPRL}.    

Ideally one wants to have the same understanding of the three charged 
system as well. The first steps in this direction were made recently in 
\cite{MthSS,MthCnn}, where some excitations of the D1--D5 systems were 
considered. These excitations corresponded to putting a quantum with 
nonzero momentum on a D1--D5 system, thus producing the system with 
two large charges and one small charge. Unfortunately for the ansatz 
taken in \cite{MthSS} the equations were too complicated to get a 
solution 
even on the linearized level, but the authors of \cite{MthSS} made an 
extensive use of matching techniques to show convincingly that the 
solution of such linearized equations should exist and it should be 
regular.  

In a seemingly unrelated line of development, recently there was a 
significant progress in understanding the properties of supergravity 
solutions in various dimensions with various amounts of supersymmetry
\cite{gmr,gmrOther}. In particular, in \cite{gmr} a classification 
of all supersymmetric solutions of minimal six dimensional supergravity 
was performed, and as we will see these are the solutions which are 
relevant for understanding of the microstates of three charged black 
hole. To be more precise, the classification of \cite{gmr} did not give 
an explicit form of a general supersymmetric solution, and the examples 
considered in \cite{gmr} had horizons and singularities, but \cite{gmr} 
derived a set of (nonlinear partial differential) equations which should 
be satisfied by all supersymmetric solutions. While we are not aware of 
any method of solving such PDEs, we will see that intuition from D1--D5 
system can be used to construct a particular solution which has three 
charges and which is completely regular. 

Our construction is inspired by the work of \cite{bal,mm}, where the 
first and the simplest example of regular geometry for D1--D5 system was 
presented. Let us briefly review their arguments. According to AdS/CFT 
correspondence \cite{mald,gkpw}, near horizon limit of D1--D5 system is 
dual to a state in two dimensional CFT. However if we start from 
asymptotically flat geometry, fermions are necessarily periodic as we 
go around the spacial direction in CFT (i.e. they belong to the Ramond 
sector). On the other hand, the vacuum of CFT belongs to NS sector, so 
it has antiperiodic fermions. In two dimensional conformal field theory 
one usually goes between R and NS sector by a procedure known as spectral 
flow. In \cite{bal,mm} it was shown that the spectral flow in CFT 
corresponds to a diffeomorphism on the gravity side. Moreover, by 
applying a spectral flow to the NS vacuum, \cite{bal,mm} produced a 
geometry which represented a near horizon limit of the black hole of 
\cite{cvet}, and thus it could be continued all the way to flat infinity.
The CFT interpretation of the near horizon limit led to a particular 
relation between the parameters of the solution, and for this set of 
parameters 
\cite{bal,mm} demonstrated that the solution was regular.
While we are not planning to discuss the general relation between 
spectral flow and gravity solutions (we just refer to 
\cite{bal,mm,MathFl}), we will apply an additional spectral 
flow to the solution of \cite{bal,mm}. This would allow us to construct a 
regular solution with three charges. It is less trivial than it sounds 
because the spectral flow would only give us solution in the near horizon 
region, and then we will have to use the technology of \cite{gmr} to find 
the interpolating functions which make the solution asymptotically flat 
rather than $AdS_3\times S^3$. 

In order to describe a microscopic state contributing to the entropy 
of D1--D5--P black hole, the geometry should satisfy the following 
requirements\footnote{The most nontrivial requirements are 2 and 3 and 
they were spelled out in \cite{MthSS}.}:
\begin{enumerate}
\item{The solution is asymptotically flat.}
\item{The solution does not have curvature singularities.}
\item{The solution does not have a horizon.}
\item{The solution has three charges.}
\item{The solution is supersymmetric.}
\end{enumerate}  
We will demonstrate that our solution possesses all these properties, 
and thus it is a good candidate for a microscopic state of a three 
charged black hole\footnote{We should mention however that for generic 
values 
of parameters the angular momenta of our solution are not the same as the 
angular momenta of BMPV black hole \cite{bmpv}, so one should not view 
our state as a {\it typical} state contributing to the entropy of the 
black hole.}.  
 
The paper has the following structure. In section 2 we review a 
construction of \cite{gmr} and we show that the regular solutions 
of \cite{lmPar} can be embedded in this construction when two 
formalisms overlap. In section 3 we introduce a map which relates 
two solutions of equations coming from \cite{gmr}: a solution 
with flat asymptotics is related to a solution which asymptotes to 
$AdS_3\times S^3$. We call this a ``near horizon map'' and we show that 
it is invertible. In section 4 we use the spectral flow and the 
near horizon map in order to construct an asymptotically flat space with 
three charges. Section 5 is devoted to the analysis of that solution, 
in particular we show that it satisfies the requirements 1--5 imposed 
on the 
microscopic state. Finally we discuss the relation of our geometry to
other approaches to the three charged system, and we also make some 
comments about 
possible applications to AdS/CFT.

\section{Six dimensional supergravity and D1--D5 system.}

In this section we will briefly summarize some of the results of 
\cite{gmr} which will be useful for our discussion.

In \cite{gmr}, Gutowski, Martelli and Reall studied supersymmetric 
solutions of six dimensional supergravity, and remarkably they found the 
most general form of such solutions. The theory under consideration 
consisted of graviton, two--form field $B_{\mu\nu}$ and a symplectic 
Majorana--Weyl gravitino $\psi^A_\mu$. The equations of motion for the 
bosonic fields are \cite{nish,gmr}:
\bea
&&G=dB,\qquad G=^\star_6 G,\nonumber\\
&&R_{\mu\nu}=G_{\mu\rho\sigma}{G_\nu}^{\rho\sigma}
\eea
The authors of \cite{gmr} were interested in constructing solutions which 
had at least one Killing spinor satisfying
\bea
\nabla_\mu \eps-\frac{1}{4}G_{\mu\nu\la}\gamma^{\nu\la}\eps=0.
\eea
Such solutions would have at least four supersymmetries. By studying 
restrictions which are imposed by the mere existence of the Killing 
spinor, GMR were able to show that the metric of the supersymmetric 
solution can always be written in the form\footnote{Here and below we are 
slightly modifying the notation of \cite{gmr} to make it simpler for the 
applications which we have in mind, but one can easily trace our notation 
to the original notation of \cite{gmr}}:
\bea\label{GMRMetr}
ds^2&=&-2H^{-1}(du+\beta_mdx^m)\left(dv+\omega_m dx^m+
\frac{F}{2}(du+\beta_mdx^m)\right)+Hh_{mn} dx^mdx^m
\eea
Here $h_{mn}$ is a hyper--K\"ahler metric of the base space. We assume 
that this metric as well as all other functions entering the ansatz 
are functions of $x_m$ only\footnote{Generically 
$h_{mn},\ \beta_m,\ \omega_m,\ H$ and $F$ 
can also depend upon the coordinate $u$, see \cite{gmr} for details.}.
Under this assumption the equations of \cite{gmr} simplify considerably, 
and we summarize them here:
\bea\label{GMROne}
&&d\beta=^\star d\beta\nonumber\\
&&d^\star dH+d\beta\wedge {\cal G}^+=0\\
&&d{\cal G}^+=0\nonumber\\
&&^\star d^\star dF=({\cal G}^+)_{mn}({\cal G}^+)^{mn}
\nonumber
\eea
Here and below an expression $^\star A$ means taking the Hodge dual with 
respect to {\it four dimensional base space} with metric $h_{mn}$, and 
the superscript $^\pm$ denotes (anti)self--dual part of a two--form under 
this Hodge duality. We also introduce the following notation:
\bea\label{GMRTwo}
&&{\cal G}^+\equiv H^{-1}((d\omega)^++\frac{1}{2}Fd\beta)
\eea
For completeness we also give an expression for the tree--form field 
strength \cite{gmr}:
\bea\label{GMRThree}
G&=&\frac{1}{2}~^\star dH-
\frac{1}{2}H^{-1}(du+\beta)\wedge (d\omega)^{-}\nonumber\\
&+&
\frac{1}{2}H^{-1}\left[dv+\omega+
\frac{F}{2}(du+\beta)\right]\wedge 
\left[d\beta+(du+\beta)\wedge dH\right]
\eea

In \cite{gmr} it was shown that any solution of the system 
(\ref{GMROne})--(\ref{GMRThree}) gives a supersymmetric solution of the 
minimal six dimensional supergravity. 
These solutions can be easily embedded in type IIB supergravity 
\cite{10People} 
by treating $B_{\mu\nu}$ as NS--NS (or R--R) two 
form and treating the metric (\ref{GMROne}) as a 
six--dimensional part of the string metric:
\bea
ds_{10}^2=ds_6^2+dz_1^2+dz_2^2+dz_3^2+dz_4^2
\eea 
Since neither dilaton, nor moduli controlling the size of four dimensional
torus are excited, we don't have to distinguish between string and 
Einstein frames. Depending on the type of the two form (NS--NS or R--R) 
the configuration describes either a system of fundamental strings 
and NS5 
branes, or a system of D1 and D5 branes. For concreteness we will 
always be talking about D1--D5 system.

Unfortunately, most of the solutions of (\ref{GMROne})--(\ref{GMRTwo})
have curvature singularities, but there exists a large class of the 
solutions which are completely regular. These solutions were first 
constructed in \cite{lmPar}, and their smoothness was proven in 
\cite{lmm}. While the regular solutions of \cite{lmPar} generically lie 
outside the scope of minimal supergravity (for example, they have 
nontrivial dilaton), there is an overlap between solutions of \cite{gmr} 
and regular solutions of \cite{lmPar}. Let us briefly summarize the 
properties of this subclass.

In \cite{lmPar} the solutions for D1--D5 system were parameterized in 
terms of a four dimensional vector ${\bf F}(\xi)$ (which was a function 
of one variable $\xi$) and a single 
charge $Q$.
As we mentioned, generically these solutions do not reduce to the 
solutions of minimal six dimensional supergravity, but they do reduce to 
the solutions of \cite{gmr} if the vector satisfies the condition:
\bea\label{NormF}
{\dot{\bf F}}^2(\xi)=0
\eea   
Using this profile one constructs a harmonic function $H$ and a gauge 
field $A_i$:
\bea\label{OurHarmProf}
H=1+\frac{Q}{L}\int _0^L\frac{d\xi}{[{\bf x}-{\bf F}(\xi)]^2},\qquad
A_i=-\frac{Q}{L}\int _0^L
\frac{{\dot F}_i(\xi)d\xi}{[{\bf x}-{\bf F}(\xi)]^2},
\eea
and constructs the field $B_i$ dual to $A_i$ with respect to a {\it flat} 
four dimensional metric:
\bea
dB=-^\star~ dA.
\eea
The regular solution of the D1--D5 system can then be constructed in 
terms of these data:
\bea\label{OurD1D5}
ds^2&=&H^{-1}\left[-(dt-A_m dx^m)^2+(dy+B_m dx^m)^2\right]+
H \delta_{mn}dx^mdx^n+
d{\bf z}d{\bf z}\\
G^{(3)}&=&d\left\{H^{-1}(dt-A_m dx^m)\wedge (dy+B_m dx^m)\right\}-
^\star dH\nonumber
\eea
To compare this with GMR solution (\ref{GMROne}), (\ref{GMRThree}), we 
introduce the light--like coordinates $u,v$ as well as self--dual and 
anti--self--dual fields $a_m$, $b_m$:
\bea\label{DefNullCrd}
u=\frac{t+y}{\sqrt{2}},\quad v=\frac{t-y}{\sqrt{2}},\qquad
a_m=\frac{B_m-A_m}{\sqrt{2}},\quad 
b_m=-\frac{A_m+B_m}{\sqrt{2}}
\eea
Using this notation we can rewrite (\ref{OurD1D5}) as
\bea\label{D1D5forGMR}
ds^2&=&-2H^{-1}(du+a_m dx^m)(dv+b_m dx^m)+
H \delta_{mn}dx^mdx^n+
d{\bf z}d{\bf z}\\
\label{MyG3}
G^{(3)}&=&d\left\{H^{-1}(dv+b_m dx^m)\wedge (du+a_m dx^m)\right\}-
^\star dH
\eea
We observe the perfect agreement between this solution and 
(\ref{GMROne}), (\ref{GMRThree}) in the region where they 
overlap\footnote{There is an overall factor of $-2$ between 
(\ref{GMRThree}) and (\ref{MyG3}), which can be traced to different 
normalization for the two forms used in \cite{lmPar} and \cite{gmr}.}, 
i.e.
\bea 
F=0,\quad \beta=a_m dx^m,\quad \omega=b_m dx^m
\eea
and thus $(d\omega)^{-}=d\omega$.

To summarize, so far we have discussed two different sets of solutions of 
type IIB supergravity reduced to six dimensions. One class \cite{gmr} 
gives solutions which have self--dual field strength in six dimensions, 
and upon lifting to 10d it describes supersymmetric solutions which 
generically have three charges: D1, D5 and momentum. Unfortunately 
solutions of this class are not guaranteed to be regular and we do not 
know any simple way to produce regular solutions in this approach.

Another class is represented by solutions of \cite{lmPar}, they all have 
regular geometries and generically six dimensional field strength is not 
self--dual and dilaton is excited. Unfortunately, those solutions have 
only two charges (D1 and D5) since in CFT they correspond to the 
Ramond vacua 
\bea
L_0={\bar L}_0=\frac{c}{24}
\eea 
so the momentum charge is necessarily zero. We also saw that in the 
region where both formalisms can be applied (i.e. for solutions without 
dilaton and without momentum charge) they give the same results. 

In the next section we will try to combine the virtues of two approaches:
microscopic intuition (which was ultimately responsible for regularity of 
the solutions in \cite{lmPar}) and the power of supersymmetry (which was 
responsible for the completely general ansatz of \cite{gmr}) to construct 
an example of a regular solution which has three charges (D1, D5 and 
momentum). 

\section{Regular solutions: far away and up--close.}

We are interested in getting solutions of (\ref{GMROne})--
(\ref{GMRThree}) which are regular and which have flat asymptotics. 
Specifically we will require that the base metric $h_{mn}$ is 
asymptotically flat,
the function $F$ as well as one forms $\beta$, $\omega$ die off at 
infinity, while $H$ can be represented as
\bea
H=1+{\hat H}
\eea 
where ${\hat H}$ goes to zero at infinity. This representation of $H$ is 
particularly useful for taking the near horizon limit \cite{mald}.

Let us start from the D1--D5 solutions (\ref{OurD1D5}). Then one 
goes to the near horizon region by replacing $H$ by ${\hat H}$. 
Since equations (\ref{GMRTwo}) in this case reduce to
\bea
d\beta=^\star d\beta,\qquad d^\star dH=0,\qquad d\omega=-^\star d\omega
\eea
it is clear that the near horizon region defined in such way is a 
solution 
of the same equations as the original system. Of course, the near horizon 
solution is no longer asymptotically flat, it asymptotes to 
$AdS_3\times S^3$, and this fact will be important in the next section.

Let us now try to define an analog of the near horizon region for the 
generic solution of (\ref{GMROne})--(\ref{GMRTwo}). We still want to 
have $H\rightarrow {\hat H}$ in this region, but now this would not be 
enough because this replacement modifies the expression for ${\cal G}^+$ 
and the equations (\ref{GMRTwo}) will no longer be satisfied. The simplest
way to resolve this problem is to modify $\omega$ as well in such a way 
that ${\cal G}^+$ does not change. More explicitly, we will define the 
``near horizon region'' as a solution obtained by a replacement
\bea
H=1+{\hat H}\rightarrow {\hat H},\qquad \omega\rightarrow {\hat\omega}
\eea  
such that
\bea
H^{-1}((d\omega)^++\frac{1}{2}F d\beta)=
{\hat H}^{-1}((d{\hat\omega})^++\frac{1}{2}F d\beta).
\eea
In other words, 
\bea\label{FlatToAdS}
{\hat H}=H-1,\qquad
(d{\hat\omega})^+=\frac{H-1}{H} (d\omega)^+ -\frac{1}{2}\frac{F}{H}d\beta
\eea
Notice that so far the ``near horizon region'' was just a name for the 
procedure which maps one solution of (\ref{GMROne})--(\ref{GMRTwo}) 
into another one (but with different asymptotics). 
Going to this region was not associated with taking any kind of a limit,
this fact allows us to invert the map:
\bea\label{AdSToFlat}
H={\hat H}+1,\qquad
(d{\omega})^+=\frac{{\hat H}+1}{{\hat H}} 
(d{\hat\omega})^+ +\frac{1}{2}\frac{F}{{\hat H}}d\beta
\eea

We arrived at the main observation of this paper. Given an asymptotically 
flat solution parameterized by $(H,F,\beta,\omega,h_{mn})$ we can 
construct a solution with $AdS_3\times S^3$ asymptotics which is 
parameterized by $({\hat H},F,\beta,{\hat\omega},h_{mn})$ by applying the 
map (\ref{FlatToAdS}). And vice versa, starting from solution which 
asymptotes to $AdS_3\times S^3$, we can construct an asymptotically flat 
solution using the map (\ref{AdSToFlat}). These maps give a one--to--one 
correspondence between solutions with flat and $AdS_3\times S^3$ 
asymptotics if we also require that $\omega$ decays at infinity.

Notice that the term ``near horizon region'' was inspired by the term 
``near horizon limit'' used for the black holes \cite{mald}. Of course, 
in our case we hope to avoid horizons altogether (and we will show that 
this indeed happens for the explicit solution), but it is still 
convenient to give some special name to the solution 
$({\hat H},F,\beta,{\hat\omega},h_{mn})$ and we choose this name to be 
``near horizon region.'' We hope that this would not lead to confusion 
and we want to stress again that the term has nothing to do with horizon.
It is also interesting that the map (\ref{FlatToAdS}) can sometimes be 
understood as a limit which is analogous to the near horizon limit for 
the black holes and we discuss this limit in more detail in the 
Appendix \ref{AppLimit}. 

Now we can proceed in constructing the solution for D1--D5 system with 
momentum charge. We will use the following strategy. First we take one of 
the regular solutions from the family (\ref{OurD1D5}) and go to the near 
horizon region using (\ref{FlatToAdS}). Then we apply a diffeomorphism 
in $AdS_3\times S^3$ to introduce a nontrivial function $F$ 
to the system. Notice that after such diffeomorphism the metric would not 
generically be in the form (\ref{GMROne}), so the inverse transformation 
(\ref{AdSToFlat}) cannot be applied\footnote{Notice that in \cite{gmr} 
is was proven that all supersymmetric solutions {\it can} be represented 
in the form (\ref{GMROne}) {\it after} appropriate change of variables. 
If after diffeomorphism the solution is not in the form (\ref{GMROne}), 
this simply means that we chose a bad coordinate system.
}. However there is a set of special diffeomorphisms (which are called 
spectral flow from the CFT point of view) which preserve the structure of 
GMR ansatz. To be more precise, a generic spectral flow gives a solution 
of \cite{gmr} which is $u$--dependent. The solutions produced by such 
spectral flows will be analyzed elsewhere, but in the next section we 
consider 
the simplest example of the spectral flow which transforms a static 
solution of the form (\ref{GMROne}) into another static solution which 
has the same form. The new solution however has a nontrivial function $F$,
so after applying the inverse transformation (\ref{AdSToFlat}) to it we 
produce an asymptotically flat solution with nonzero value of the 
momentum. Since this solution was regular in the ``near horizon'' region 
it is plausible that it will be regular everywhere. 

Let us summarize our strategy.
\begin{enumerate}
\item{Start from one of the regular solutions (\ref{OurD1D5}) and rewrite 
it in the form (\ref{D1D5forGMR})
}
\item{Go to the ``near horizon region'' using (\ref{FlatToAdS}).}
\item{Perform a spectral flow which keeps the solution in the class 
(\ref{GMROne}), but produces nontrivial $F$.}
\item{Change boundary conditions to flat by using (\ref{AdSToFlat}).}
\end{enumerate}

In the end we produce a solution which has three charges and which has 
a good chance to be 
completely regular. We will present an explicit example of this 
construction in the next section. 
  
\section{An example of a solution with three charges.}

Let us now implement the general procedure which was outlined in the 
previous section. We begin with solution describing extremal two charged 
black hole which was considered in \cite{bal,mm}\footnote{This solution 
was a special case of a more general five dimensional black holes 
constructed in \cite{cvet}.}. In \cite{bal,mm} it was observed that 
for certain values of the parameters the 
near horizon region of such black hole becomes $AdS_3\times S^3$ in 
{\it global} coordinates. This regular solution can be viewed as a special
case of the general metric (\ref{OurD1D5}) with a particular 
profile\footnote{
This way of deriving solution from chiral null model \cite{cnm} 
was presented in \cite{Multi}}:
\bea
F_1(\xi)=a\cos\frac{2\pi\xi}{L},\qquad F_2(\xi)=a\sin\frac{2\pi\xi}{L},
\qquad F_3(\xi)=F_4(\xi)=0.
\eea
Notice that while this solution is regular for all values of $a$ and $L$,
generically it falls outside the scope of minimal supergravity (for 
example, generically it has a nontrivial dilaton). As we discussed before,
the solution belongs to minimal supergravity if and only if
\bea
|{\dot{\bf F}}|=1\quad\rightarrow\quad L=2\pi a.
\eea
From now on we will assume that this relation is satisfied.

We present the complete solution later (this solution will 
come out as a special member of a more general class), and here we 
directly write the 
near horizon geometry, i.e. 
we go directly to the second step in our strategy:
\bea\label{MMNearHor}
ds^2&=&\frac{1}{Q}\left[-(r^2+a^2)dt^2+r^2 dy^2\right]+
Q\frac{dr^2}{r^2+a^2}
\nonumber\\
&+&
Q\left[d\theta^2+\cos^2\theta(d\psi-\frac{a}{Q}dy)^2
+\sin^2\theta(d\phi-\frac{a}{Q}dt)^2\right]
\eea 
Notice that we wrote this metric in a form which is slightly different 
from (\ref{GMRMetr}), but which is equivalent to it: one should 
use the expressions for the 
null coordinates (\ref{GMRMetr}) and recombine various terms. The reason 
we chose
this ``unconventional'' form is that it makes the $AdS_3\times S^3$ 
structure
explicit and also it makes obvious the fact that metric is regular if the
radius\footnote{We use identification $y\sim y+2\pi R$.} 
of $y$ circle takes a special value \cite{bal,mm}:
\bea\label{RegRadNrHor}
R=\frac{Q}{a}
\eea 
Notice that the bosonic metric (\ref{MMNearHor}) describes 
$AdS_3\times S^3$ in global coordinates, but due to nontrivial cross 
terms between sphere and AdS, the fermions are {\it periodic} under 
identification $y\sim y+2\pi R$, i.e. we are dealing with 
{\it Ramond sector} of the corresponding CFT\footnote{This fact is not 
surprising, since (\ref{MMNearHor}) was obtained as a near horizon limit 
of asymptotically flat geometry, where fermions were clearly periodic.}.
To see the relation between NS and R sectors more clearly, we rewrite the
metric in terms of the coordinates $u$ and $v$:
\bea\label{BeforeSF}
ds^2&=&-\frac{1}{Q}\left[2r^2 du dv+\frac{Q^2}{2R^2}(du+dv)^2\right]+
Q\frac{dr^2}{r^2+Q^2/R^2}
\nonumber\\
&+&
Q\left[d\theta^2+\cos^2\theta\left(d\psi-\frac{du}{\sqrt{2}R}+
\frac{dv}{\sqrt{2}R}\right)^2
+\sin^2\theta\left(d\phi-\frac{du}{\sqrt{2}R}-
\frac{dv}{\sqrt{2}R}\right)^2\right]
\eea
The connections on the sphere which appear in the second line of the 
above expression are responsible for performing the spectral flow from 
the NS vacuum of CFT. In particular, the connections proportional to 
$du$ are responsible to the spectral flow in the left sector and the ones 
proportional to $dv$ correspond to spectral flow in the right sector (see
\cite{bal,mm,MathFl} for details). An addition of extra connection 
proportional to $du$ performs an extra spectral flow in the left sector, 
while still keeping Ramond vacuum on the right:
\bea
{\bar L}_0=\frac{c}{24}
\eea    
Let us perform such spectral flow:
\bea\label{AfterSF}
ds^2&=&-\frac{1}{Q}\left[2r^2 du dv+\frac{Q^2}{2R^2}(du+dv)^2\right]+
Q\frac{dr^2}{r^2+Q^2/R^2}+Qd\theta^2
\\
&+&
Q\left[\cos^2\theta\left(d\psi-(2\nu+1)\frac{du}{\sqrt{2}R}+
\frac{dv}{\sqrt{2}R}\right)^2
+\sin^2\theta\left(d\phi-(2\nu+1)\frac{du}{\sqrt{2}R}-
\frac{dv}{\sqrt{2}R}\right)^2\right]\nonumber
\eea
In this expression $\nu$ is just a parameter which describes the extra 
spectral flow. However we will be interested in the case where the new 
geometry is regular, then $\nu$ has to be an integer. To see this we 
observe that the metrics (\ref{BeforeSF}) and (\ref{AfterSF})  
are related by diffeomorphism:
\bea
\psi\rightarrow \psi-2\nu\frac{u}{\sqrt{2}R},\qquad
\phi\rightarrow \phi-2\nu\frac{u}{\sqrt{2}R},
\eea   
and such diffeomorphisms relating periodic variables (we recall that $u$ 
inherits periodicity from $y$ coordinate: $u\sim u+\sqrt{2}\pi R$) are 
regular only if $\nu$ is integer\footnote{Another reason to concentrate on
integer $\nu$ comes from CFT: since we want to flow from one state in 
the Ramond sector to another state in the same sector, then $\nu$ has to 
be 
an integer. The detailed discussion of relation between spectral flow 
and supergravity solutions is beyond the scope of this paper.}.

At this point we almost completed step 3 of our program: we found a 
solution
in the near horizon region which has nontrivial $F$ (i.e. it has 
nontrivial $g_{uu}$). To be able to extend this solution to the 
asymptotically flat region, we should recombine it into the form 
(\ref{GMROne}). Performing a simple algebra, we find the functions 
parameterizing the solution:
\bea\label{NewHhat}
{\hat H}=\frac{Q}{r^2+(\nu+1)a^2\cos^2\theta-\nu a^2\sin^2\theta},\quad
F=-\frac{2\nu(\nu+1)a^2}{r^2+(\nu+1)a^2\cos^2\theta-\nu a^2\sin^2\theta}
\eea     
\bea
\beta=\frac{aQ}{\sqrt{2}}\frac{\sin^2\theta d\phi-\cos^2\theta d\psi}{
r^2+(\nu+1)a^2\cos^2\theta-\nu a^2\sin^2\theta}
\eea
\bea\label{NewOmhat}
{\hat\omega}&=&\frac{aQ}{\sqrt{2}}
\frac{(1+2\nu)r^2+\nu a^2\sin^2\theta+(\nu+1)(1+4\nu)a^2\cos^2\theta}
{(r^2+(\nu+1)a^2\cos^2\theta-\nu a^2\sin^2\theta)^2}\sin^2\theta d\phi
\nonumber\\
&+&\frac{aQ}{\sqrt{2}}
\frac{(1+2\nu)r^2+(\nu+1)a^2\cos^2\theta-\nu(3+4\nu)a^2\sin^2\theta}
{(r^2+(\nu+1)a^2\cos^2\theta-\nu a^2\sin^2\theta)^2}\cos^2\theta d\psi,
\eea
as well as the metric of the base space:
\bea\label{NewBase}
&&h_{mn}dx^m dx^n=f
\left[
\frac{dr^2}{r^2+a^2}+d\theta^2
\right]
-\frac{a^4}{2f}\nu(\nu+1)\sin^2 2\theta d\phi d\psi
\nonumber\\
&&\quad+\frac{1}{f}
\left\{
(r^2+a^2\cos^2\theta)(r^2+a^2)+\nu a^2\cos^2\theta(2r^2+(\nu+2)a^2)
\right\}\sin^2\theta d\phi^2\nonumber\\
&&\quad+\frac{1}{f}
\left\{
(r^2+a^2\cos^2\theta)r^2+\nu a^2\sin^2\theta(-2r^2+\nu a^2)
\right\}\cos^2\theta d\psi^2
\eea
To simplify the expressions we introduced a convenient notation:
\bea
f\equiv r^2+(\nu+1)a^2\cos^2\theta-\nu a^2\sin^2\theta.
\eea
We also traded the radius of $y$ direction for the 
parameter $a$:
\bea\label{RtoANearHor}
a=\frac{Q}{R}
\eea
and from now on the radius $R$ will not appear in the solution. 
Jumping ahead, we just mention that the relation (\ref{RtoANearHor}) 
holds for the near horizon solution, but it will look slightly 
different for the one which is asymptotically flat.   

In equations (\ref{NewHhat}) and (\ref{NewOmhat}) we also put the hats 
over $H$ and $\omega$ to stress the fact that we are dealing with near 
horizon region.

Now we are ready to perform the final step of our program: to go from 
the space with $AdS_3\times S^3$ asymptotics to the asymptotically flat 
space by performing the map (\ref{AdSToFlat}). Let us introduce 
\bea
{\tilde\omega}=\omega-{\hat\omega}
\eea
Then (\ref{AdSToFlat}) gives an equation
\bea
(d{\tilde\omega})^+=\frac{1}{{\hat H}} 
(d{\hat\omega})^+ +\frac{1}{2}\frac{F}{{\hat H}}d\beta
\eea
This equation can be solved by taking 
\bea
{\tilde\omega}=\sqrt{2}\nu(\nu+1)a^3
\frac{\sin^2\theta d\phi-\cos^2\theta d\psi}{
r^2+(\nu+1)a^2\cos^2\theta-\nu a^2\sin^2\theta}=
\frac{2\nu(\nu+1)a^2}{Q}\beta
\eea

To summarize, we found an asymptotically flat solution of type IIB 
supergravity which has a form
\bea\label{MySlnMetric}
ds^2&=&-2H^{-1}(du+\beta_mdx^m)\left(dv+\omega_m dx^m+
\frac{F}{2}(du+\beta_mdx^m)\right)\nonumber\\
&+&Hh_{mn} dx^mdx^m+\sum_{i=1}^4 dz_idz_i\nonumber\\
G&=&\frac{1}{2}~^\star dH-
\frac{1}{2}H^{-1}(du+\beta)\wedge (d\omega)^{-}\\
&+&
\frac{1}{2}H^{-1}\left[dv+\omega+
\frac{F}{2}(du+\beta)\right]\wedge 
\left[d\beta+(du+\beta)\wedge dH\right]\nonumber
\eea
with coefficients given by (\ref{NewHhat})--(\ref{NewBase}) and
\bea\label{MySlnFlat}
H=1+{\hat H},\qquad \omega={\hat\omega}+\frac{2\nu(\nu+1)a^2}{Q}\beta
\eea
We claim that this solution is completely regular and it carries three 
nontrivial charges: D1, D5 and momentum\footnote{Notice that taking 
$\nu=0$ we recover the asymptotically flat solution of \cite{bal,mm} in 
the form it was presented in \cite{Multi}. Of course, the case $\nu=0$ is 
special, since the solution has only two charges (D1 and D5).}. 

\section{Properties of the solution.}
\label{SectProp}

In this section we will analyze the solution in more detail and we will 
show that it possesses all the right properties to be one of the 
microscopic states which contribute to the entropy of the three charged 
black hole.  

{\bf 1. The solution is asymptotically flat.}

{\bf 2. The solution does not have curvature singularities.} A sufficient 
condition for the absence of curvature singularities is to have a 
well--defined metric and inverse metric. A potential problem with 
metric $g_{\mu\nu}$ may only arise when one of the coefficient functions 
blows up. This could happen only in the regions where $f=0$ or $H=0$. 
To avoid complications associated with $H=0$, we restrict our attention 
to the range of parameters in which $H$ never vanishes\footnote{This 
does not mean however that $H$ has a definite sign. But the regions 
with positive and negative $H$ are connected through $H=\infty$, 
not $H=0$.}:
\bea\label{RangeForH}
\nu>0~~ \Rightarrow~~ a^2<\frac{Q}{\nu},\qquad{\mbox{or}}\qquad
\nu<0~~ \Rightarrow~~ a^2<\frac{Q}{|\nu+1|}
\eea
While it would be interesting to understand the properties of the 
solution outside the range (\ref{RangeForH}), our goal is to present 
the simplest example of the regular solution, so we will assume that 
(\ref{RangeForH}) is satisfied. Under this assumption, particular 
components of the metric $g_{\mu\nu}$ can only diverge if $f=0$.

In the regions where the metric 
$g_{\mu\nu}$ is regular, the inverse metric is well--defined as long as 
$g_{\mu\nu}$ has non--vanishing determinant. By direct computation one 
can check that
\bea\label{detmet}
{\mbox{det}}~g=-\frac{1}{4}H^2 r^2\sin^2 2\theta f^2,
\eea
so the inverse metric and curvature invariants can only have problems 
if at least one of the conditions
\bea\label{ListBadPoints}
r=0,\qquad \theta=0,\qquad \theta=\frac{\pi}{2},\qquad f=0
\eea
is satisfied. Let us analyze these regions one by one.

(a) We first show that points where $\theta=0$ or $\theta=\frac{\pi}{2}$ 
correspond to coordinate singularities. To see this we go from spherical 
coordinates $(r,\theta,\psi,\phi)$ to Cartesian coordinates in the 
standard way:
\bea
&&x_1=r\sin\theta\cos\phi,\qquad x_2=r\sin\theta\sin\phi\nonumber\\
&&x_3=r\cos\theta\cos\psi,\qquad x_2=r\cos\theta\sin\psi\nonumber
\eea
Then various components of the metric may acquire additional singularities 
as $r$ goes to zero, but not at the points where $f\ne 0$ and $r\ne 0$. 
At such points we can compute the determinant of the new metric by 
multiplying (\ref{detmet}) and appropriate Jacobian:
\bea
{\mbox{det}}~g'=-\frac{1}{r^4}H^2f^2.
\eea
This demonstrates that in Cartesian coordinates both $g_{\mu\nu}$ and 
inverse metric are well--defined away from the points where $f=0$ or 
$r=0$, so the singularities at $\theta=0,\frac{\pi}{2}$ are the usual 
coordinate singularities of the spherical frame.

(b) We now look at the vicinity of the region where $r=0$. One can write 
an approximate expression for the metric of the base space:
\bea\label{ApproxBase}
h_{mn}dx^m dx^n&\approx& f
\left[
\frac{dr^2}{a^2}+d\theta^2
\right]+\frac{a^2}{f}(a^2\sin^2\theta\cos^2\theta+r^2) 
[(\nu+1)d\phi-\nu d\psi]^2
\nonumber\\
&-&\frac{a^2}{f}
(\cos 2\theta+\frac{1+2\nu}{4}(3+\cos 4\theta))r^2
[(\nu+1)d\phi-\nu d\psi][d\phi-d\psi]\nonumber\\
&+&\frac{a^2}{f}((1+\nu)\cos^2\theta-\nu\sin^2\theta)^2 r^2
[d\phi-d\psi]^2
\eea
An explicit form of this metric is not important for us, what is 
important is that this metric is regular in the new coordinate system:
\bea\label{RegREq0}
{\tilde\phi}=(\nu+1)\phi-\nu\psi,\quad {\tilde\psi}=\psi-\phi,\qquad
x_1=r\cos{\tilde\psi},\quad x_2=r\sin{\tilde\psi}.
\eea
In this coordinate system the determinant (\ref{detmet}) gets 
multiplied by the appropriate Jacobian to become
\bea
{\mbox{det}}~g'\sim H^2 \sin^2 2\theta f^2,
\eea 
so both direct and inverse metrics are regular at $r=0$ unless $f=0$ as 
well. To be more precise, the introduction of Cartesian coordinates may 
lead to new singularities in metric components if $d{\tilde\psi}$ appears
without $r^2$ in front of it. However, one can check that such 
singularities can be eliminated by an additional diffeomorphism
\bea\label{R0Twist}
u\rightarrow {\tilde u}=u-\frac{Q}{\sqrt{2} a}{\tilde\psi},\qquad
v\rightarrow {\tilde v}=v+\frac{Q}{\sqrt{2} a}\left[1-
2\nu(\nu+1)\frac{a^2}{Q}\right]{\tilde\psi},
\eea  
We present the details in the Appendix \ref{AppSingul}. This completes 
the proof of 
regularity at $r=0$ and generic value of $\theta$.

We still have 
to show that metrics are regular if both $r=0$ and $\theta=0$ 
(or $r=0$ and $\theta=\pi/2$). One can check that all curvature 
invariants stay finite as we approach such points 
(we present some details in the Appendix \ref{AppSingul}).
However for generic values of the parameters, the solution develops 
a conical singularity at $r=0$ and poles on the sphere. The singularity 
is absent if and only if the parameters satisfy the relations 
(\ref{RegCondApp})\footnote{From the point of view of an observer at 
infinity the first of the relations (\ref{RegCond}) should be interpreted 
backward. One can fix the charge of the solution $Q$, the radius of 
$y$ direction $R$ and the integer $m$, then (\ref{RegCond}) determines 
$a$ and $\nu$. Notice that $\nu$ is not necessarily an integer.}:
\bea\label{RegCond}
R=\frac{Q}{a}|m-1|,\qquad \nu(\nu+1)=\frac{Qm}{a^2}
\eea
with some integer $m$. Notice that for $m=0$ this reduces to the familiar 
regularity condition for D1--D5 (\ref{RegRadNrHor}):
\bea
R=\frac{Q}{a}
\eea
We also notice that relation (\ref{RegCond}) is different from the 
condition which we had in the near horizon region:
\bea
R=\frac{Q}{a},\qquad \nu=m
\eea
and generically $\nu$ is (\ref{RegCond}) is not an integer.

The source of this difference is easy to explain: the near horizon 
limit of 
asymptotically flat solution is supposed to work only in the vicinity of 
the surface $f=0$. However for $m\ne 0$ the points $r=0,\theta=0$ are not 
close to this surface, so it is not surprising that the regularity 
condition was modified.  

(c) Finally we consider a vicinity of the points where function $f$ 
vanishes. The analysis is somewhat technical, and we refer to the 
Appendix \ref{AppSingul} for the details. Here we just state that the 
solution does not have singularities at such points.

To summarize, we have shown that all apparent singularities of the 
solution (\ref{NewHhat})--(\ref{NewBase}), (\ref{MySlnMetric}), 
(\ref{MySlnFlat}) can be traced to a bad choice of 
coordinates, and all curvature invariants are well--defined and 
finite for our solution. This means that the solution is completely 
regular.

{\bf 3. The solution does not have a horizon.} To check this claim one
needs to study the global properties of geodesics, and we will not 
present such detailed analysis here. Instead we will look for one of the 
{\it symptoms} of the horizon: infinite redshift of the frequency. This
will allow us to rely on {\it local} analysis, and although it would not 
rigorously prove the absence of the horizon, it will give a strong 
argument in support of this claim.    
 
So we want to look for the surfaces of infinite redshift, i.e. surfaces 
where $g^{tt}$ blows up. As we already observed in the analysis of 
possible singularities, the components of the inverse metric (including 
$g^{tt}$) can only blow up at the points (\ref{ListBadPoints}) 
where either individual components of the metric diverge or determinant 
of the metric vanishes. Now we have to go through the same list.

(a) For $\theta=0$ or $\theta=\frac{\pi}{2}$ we again go to the Cartesian 
coordinates to see that $g^{tt}$ stays finite. 

(b) In the vicinity of the region $r=0$ the coefficient $g^{tt}$ does 
blow up, since the determinant of the metric vanishes as $r^2$, while the 
cofactor of $g^{tt}$ behaves as
\bea
A^{tt}\sim -\frac{a^4\nu^2(\nu+1)^2}{4Q}
\frac{Q+(\nu+1)a^2\cos^2\theta-\nu a^2\sin^2\theta}{
(\nu+1)\cos^2\theta-n\sin^2\theta}\sin^2 2\theta
\eea
However this surface of infinite redshift should not be interpreted as a 
horizon. The reason is that unlike the case of Schwarzschild black hole 
where coordinates $(r,t)$ break down at the horizon and one needs to use 
analytic continuation in both of them, here at the surface $r=0$ $r$ is 
still a good coordinate. Moreover, in the coordinate frame 
(\ref{RegREq0}) 
which regularizes this point, $r$ is a radial coordinate, so the space 
is complete and we cannot continue beyond $r=0$. This means that while 
the surface of infinite redshift at $r=0$ is a candidate for a horizon, 
there would be no space ``behind it,'' so we this surface should not be 
viewed as a horizon.\footnote{This should be contrasted with the case of 
conventional three charge black hole \cite{ts9601} with harmonic 
functions 
$H_i=Q_i/r^2$. In that case the horizon was also located at $r=0$, but 
the coordinate system was singular there, so one needs to do an analytic 
continuation and one indeed sees a horizon.}    

(c) In the vicinity of the points where $f=0$ the function $g^{tt}$ goes 
to a finite limit. To see this we observe that near these points the 
metric 
has a form (\ref{SumPertMetr}) with $\alpha=1$ and all its components are 
regular. 
We also know that (\ref{detmet}) goes to a finite limit as we approach 
$f=0$, so all components of $g^{\mu\nu}$ (and in particular $g^{tt}$) 
stay finite.

To summarize, we have analyzed the surfaces of infinite redshift and we 
have shown that there is only one such surface ($r=0$), but it should not 
be viewed as a horizon. It would be interesting to perform a more 
detailed study of causal structure of the space--time which we are 
considering to reach a definitive conclusion about the presence of the 
horizon. 
Such investigation should also shed some light on the closed time--like 
curves in the geometry. Here we will rely only on the local analysis, 
and we take the absence of the surfaces with infinite redshift as a 
strong indication for the absence of the horizon.   

{\bf 4. The solution has three charges.} In order to compute them we 
have to look at the fall--off of various fields at infinity. The numbers 
of D1 and D5 branes and momentum excitations are\footnote{An 
extra factor of two in these expressions appears due to non--traditional 
normalization of $G$ in \cite{gmr}} \cite{hms}:
\bea
n_1=2\frac{V}{4\pi^2 g}\int_{S^3}~ ^\star_{6} G=\frac{QV}{g},\qquad
n_5=\frac{2}{4\pi^2 g}\int_{S^3}~ G=\frac{Q}{g},\qquad
n_P=\frac{R^2 V}{g^2}\nu(\nu+1)=\frac{R^2 V}{g^2}\frac{Q}{a^2}m
\eea
For completeness we also present the expressions for the angular momenta:
\bea
J_\phi=\frac{aR}{Q}\frac{Q^2 V}{g^2}\left(
1+\nu\left[1+\frac{a^2}{Q}(\nu+1)\right]
\right),\qquad
J_\psi=\frac{aR}{Q}\frac{Q^2 V}{g^2}\nu\left[1-\frac{a^2}{Q}(\nu+1)\right]
\eea

{\bf 5. The solution is supersymmetric.} This was a starting assumption 
of the GMR construction \cite{gmr}, and our metric solves their equations.

{\bf 6. Interpretation in terms of branes.}\footnote{I want to thank 
Juan Maldacena for suggesting to add this discussion.}
It is interesting to find a configuration of branes which produces the 
solution which we constructed. According to the usual correspondence 
between branes and supergravity solutions \cite{sussk}, at weak string 
coupling one starts from branes in flat space, then as $g$ gets larger, 
the branes start to modify the geometry producing a nontrivial 
gravitational background. In this paper we constructed an example of 
such background and we showed that it is regular. Now we want to go to a 
weak string coupling and identify the corresponding configuration of 
branes. Since we want to keep charges fixed, we would be interested in 
the following rescaling:
\bea\label{BraneScale}
g\rightarrow \eps g, \qquad Q\rightarrow \eps Q,\qquad R\rightarrow \eps R
\eea      
and we keep the values of $V,a,m$ fixed. As $\eps$ goes to zero, we 
observe that 
\bea
\nu\sim \frac{Qm}{a^2}
\eea 
becomes small, and then it is convenient to write function $f$ in the 
form
\bea
f=r^2+a^2\cos^2\theta+\nu a^2\cos 2\theta
\eea

First we consider the region where 
$f\gg\nu a^2\sim mQ$,
and we always 
assume that $m\ne 0$. Then the metric 
of the 
base space (\ref{NewBase}) reduces to
\bea
&&h_{mn}dx^m dx^n=(r^2+a^2\cos^2\theta)
\left[
\frac{dr^2}{r^2+a^2}+d\theta^2
\right]+(r^2+a^2)\sin^2\theta d\phi^2+r^2\cos^2\theta d\psi^2
\eea
This is just a metric of a flat four dimensional space written in the 
unusual coordinates (see \cite{Multi} for details). One can also see 
that the functions $F,\beta,\omega$
can be dropped from (\ref{MySlnMetric}) in this region and $H$ can 
be replaced by one. So in the 
region where $f\gg mQ$ we have a usual flat space. However this condition 
itself cuts some region out of the flat space, and we will now 
describe this region. In a four dimensional space parameterized by the 
Cartesian coordinates $x_1,x_2,x_3,x_4$, we take a 1--2 plane and draw 
a circle of radius $a$ with a center in the origin. Then we consider 
a three dimensional torus which surrounds this circle and which has 
radii $(a,d,d)$ where $a\gg d\gg\sqrt{mQ}$. Notice that pictorially 
this torus looks like a thin tube 
surrounding the circle. One can see that outside 
of this torus the condition $f\gg mQ$ is satisfied and the space is 
flat.

Now we look at the interior of the torus. Before we do this for the three 
charged system, it is useful to recall the picture for the two charges. 
In that case the space was also flat outside the torus 
$(a,d',d')$ (where $a\gg d'\gg \sqrt{Q}$), and the branes themselves were 
located 
precisely on the circle $x_1^2+x_2^2=a^2,x_3=x_4=0$. Inside the torus the 
system had some curved geometry, but as one approached the circle one saw 
a metric of the KK monopole with D1 and D5 fluxes \cite{lmm}.
In the three charged system we would also have some complicated geometry 
as we go inside the torus $(a,d,d)$, but we want to see how 
this geometry ends. To analyze this one needs to consider the limit 
$f\ll Q$, which selects the following 
region in the $r,\theta$ space:
\bea
{\tilde\theta}\equiv \frac{\pi}{2}-\theta\sim
\frac{\sqrt{Q}}{a}\ll 1,\qquad 
r^2+a^2{\tilde\theta^2}\approx a^2\nu\approx mQ
\eea 
We see that unlike the case of D1--D5 system where KK monopole looked 
like a point in the $(r,{\tilde\theta})$ plane, we now have a circle 
in this 
plane\footnote{To be precise, we have a quarter of the circle since both 
$r$ and ${\tilde\theta}$ are positive.}.
Of course, the $(r,{\tilde\theta})$ plane does not have a clear 
geometrical meaning 
since these are not the Cartesian coordinates, so to analyze the shape of 
the ``singularity'' we have to consider the full six dimensional space.  
However the observation that we are dealing with a circle in 
$(r,{\tilde\theta})$ plane is still useful, because we see that at a 
generic point of this circle (when $r\ne 0$ and ${\tilde\theta}\ne 0$) 
the ``singularity'' extends along one coordinate (an angle along the 
circle) instead of two 
($r$ and ${\tilde\theta}$), and the other coordinate becomes transverse 
to the ``singularity''. On the other hand, at a generic point of 
this circle, all components of six dimensional metric are well--defined 
(see (\ref{SumPertMetr})) and its determinant (\ref{detmet}) does not 
vanish, 
so we have a non--degenerate six dimensional space. This shows that 
the worldvolume of the ``singularity'' is $1+4$ dimensional (ignoring 
the four directions on the torus), as opposed to the D1--D5 system, where 
the corresponding worldvolume had $1+1$ dimensions. Of course we don't 
actually have a singularity in either case, and it would be interesting 
to get an intuitive understanding of the regularization mechanism 
for the three charged system.   
It is also interesting to note that in the case of D1--D5 system 
the space 
ended on the singularity, while in the case of the three charges the 
space 
continues inside the ``domain wall'' and ends there smoothly. 

\section{Discussion.}

We have constructed a regular solution of type IIB supergravity which has 
three charges corresponding to D1, D5 branes and momentum. We showed that 
this solution satisfies all requirements one wants to impose on a 
microscopic state, so we conjecture that our solution would indeed be 
one of the states contributing to the entropy of three charged black 
hole\footnote{Our solutions would contribute to the entropy of either 
BMPV black hole \cite{bmpv} or more general rotating black holes such 
as ones discussed in \cite{herdeiro}.}.  
It would be very interesting to find geometries which correspond to other 
microscopic states. 

In the case of D1--D5 system the guiding principle 
which led to construction of regular geometries in \cite{lmPar} was based 
on microscopic understanding of D1--D5 bound states as being dual to a 
fundamental string vibrating with different profiles. It seems that we 
are still missing this detailed understanding of bound states with three 
charges, although a significant progress in this direction has been made 
recently \cite{DBI3Tube}. These papers tried to understand the three 
charged system in a dual frame where they describe supertube 
\cite{supertube,tubeOther}. Although the analysis of \cite{DBI3Tube} was 
done on 
the level of worldvolume theory, one may hope that ultimately there would 
be some way of finding the gravitational solutions corresponding to 
supertubes with three charges, just like the solutions for supertubes  
\cite{supertube} were found in \cite{gravTube}. 

Another interesting direction is to use the technology of \cite{gmr} in 
order to study the AdS/CFT correspondence. Again an analogy with pure 
D1--D5 system might be useful. So far we were always talking about 
solutions which were asymptotically flat. However (\ref{OurD1D5}) can 
also be viewed as solutions with $AdS_3\times S^3$ asymptotics. As such 
they would describe the geometries which are dual to Ramond vacua of the 
CFT, i.e. to the states satisfying
\bea\label{CFTRamVac}
L_0={\bar L}_0=\frac{c}{24}.
\eea
The map between such Ramond vacua and geometries was presented in 
\cite{lmPar,lmm}. The momentum charge from the point of view of CFT 
corresponds 
to $L_0-{\bar L}_0$, so in order to construct a state with nonzero 
momentum we clearly have to give up the relation (\ref{CFTRamVac}). On 
the other hand, we are still interested in supersymmetric configuration, 
so we would like to stay in the Ramond vacuum at least in one of the 
sectors. Thus the geometries with nonzero momentum would correspond to 
the states with 
\bea
L_0\ne{\bar L}_0=\frac{c}{24}.
\eea  
The near horizon region of the geometry which we discussed in this paper 
corresponded to such state, but it was not very interesting from the 
AdS/CFT point of view since it was obtained form the Ramond vacuum by a  
spectral flow. It would be very interesting to use the technology of GRM 
\cite{gmr} to construct the geometries corresponding to less trivial 
states in the CFT.

To conclude, we consider this paper as one of the first steps in 
understanding regular solutions with three charges, and getting more 
general solutions would be very important for black hole physics and for 
AdS/CFT.  

\section*{Acknowledgements.}
 
It is a pleasure to express my gratitude to Samir Mathur for numerous 
conversations which inspired this investigation. I also want to thank 
Vijay Balasubramanian, Aki Hashimoto, Juan Maldacena, Raul Rabadan and 
Joan Simon for useful discussions and David Mateos for pointing out a 
misprint in the previous version of the paper. 
This work was supported by NSF 
grant PHY--0070928.

\appendix
\section{Definitions of the near horizon limit.}
\label{AppLimit}

In this paper we reserved the term 
``near horizon region'' for the solutions parameterized 
by $({\hat H},F,\beta,{\hat\omega},h_{mn})$. Such solutions 
were defined by (\ref{FlatToAdS}) which did not involve any 
limit. In this appendix we will show that in some circumstances 
the map (\ref{FlatToAdS}) can indeed be interpreted as taking a  
limit which is analogous to the near horizon limit for the black holes
\cite{mald}. Hopefully this would give a reasonable 
justification for the term ``near horizon region.''  

We begin with the simple case of two charge solution (\ref{OurD1D5}), 
(\ref{OurHarmProf}). Let us pick some point $\xi_0\in (0,L)$ and 
zoom in on 
the region near ${\bf x}={\bf F}(\xi_0)$. Without the loss of 
generality we assume that $\xi_0=0$, ${\bf F}(0)=0$, then the 
``zooming in'' corresponds to rescaling:
\bea
{\bf x}\rightarrow \eps {\bf x}, \qquad
{\bf F}(\xi)\rightarrow \eps {\bf F}({\tilde \xi})\equiv
\eps {\bf F}(\frac{\xi}{\eps})
\eea 
Notice that we had to rescale the argument of ${\bf F}$ since 
we want to preserve the condition (\ref{NormF}). In the rescaled 
coordinates we find:
\bea
H'&=&1+\frac{Q}{\eps^2 {\tilde L}}\int_{-{\tilde L}/2}^{{\tilde L}/2}
\frac{d{\tilde \xi}}{[{\bf x}-{\bf F}({\tilde\xi})]^2}
\equiv 1+\frac{1}{\eps^2}{\hat H},\nonumber\\
A'_i&=&-\frac{Q}{\eps^2 L}\int _{-{\tilde L}/2}^{{\tilde L}/2}
\frac{{\dot F}_i({\tilde\xi})
d{\tilde\xi}}{[{\bf x}-{\bf F}({\tilde\xi})]^2}=\frac{1}{\eps^2}A_i,
\eea
In this appendix we use primes to denote expressions after rescaling.
Taking Hodge dual with respect to primed variables, we find:
\bea
B_i'=\frac{1}{\eps^2}B_i
\eea
Writing the metric (\ref{OurD1D5}) for primed variables we get:
\bea
ds^2=(\eps^2+{\hat H})^{-1}
\left[-(dt-A_m dx^m)^2+(dy+B_m dx^m)^2\right]+
(\eps^2+{\hat H}) \delta_{mn}dx^mdx^n+d{\bf z}d{\bf z}\nonumber
\eea 
Then in the limit $\eps\rightarrow 0$ we recover precisely the near 
horizon limit parameterized by $({\hat H},F,\beta,{\hat\omega},h_{mn})$.

Of course, for the profile with ${\bf F}=const$ this definition of near 
horizon limit reduces to the standard decoupling limit of \cite{mald}, 
where the role of $\eps$ was played by $\alpha'$. For this type of 
profile 
one can define a region which has exactly the same singularity, but it 
asymptotes to $AdS_3\times S^3$. 

For the profiles with 
non--constant ${\bf F}$ one should be a little more careful since there 
is one additional length scale in the problem (the characteristic size of 
$|{\bf F}(\xi_1)-{\bf F}(\xi_2)|$) which is an input parameter and 
which cannot be rescaled. So after taking $\eps$ to zero we zoom in on 
a vicinity of some point on the curve ${\bf x}={\bf F}(\xi)$, 
but generically a part of the curve would go outside this region. So 
generically there would not be a good limit which preserves the entire 
curve and has $AdS_3\times S^3$ asymptotics, but the map 
\bea
H=1+{\hat H}\rightarrow {\hat H}
\eea
can still be considered as a solution generating technique which 
produces a 
geometry with such properties. Using analogy with ${\bf F}=const$ 
case we will refer to this map as a ``going to the near horizon regime'', 
but one should keep in mind that it can't always be defined a limit. 

Now let us look at a more interesting case of solution with three charges
(\ref{GMRMetr}) and try to justify the map (\ref{FlatToAdS}) by taking 
some 
limit. From our experience with D1--D5 case we already know that this 
limit would involve ``zooming in'' on a vicinity of a point where sources 
of $H$ are located. Without the loss of generality we take this 
point to be ${\bf x}=0$. For all solutions which we consider in this 
paper (and we believe this would be generically true for the solutions 
corresponding to microscopic states), the other functions 
($\beta,\omega,F$) have sources at the same points as $H$. We 
observe that the system (\ref{GMROne}) is invariant under the 
following rescaling:
\bea\label{NearHorLimit}
x\rightarrow x'=\eps x,\quad v\rightarrow v'=\frac{v}{\eps^2},\qquad
H\rightarrow H'=1+\frac{{\hat H}}{\eps^2},
\quad \omega\rightarrow \omega'=\frac{\omega}{\eps^2},\quad
F\rightarrow F'=\frac{F}{\eps^2}
\eea   
while $u$, $\beta$ and $h_{mn}$ are kept fixed. So starting from 
any asymptotically flat solution and a point where $H$ blows up, we can 
construct a set of 
solutions which is parameterized by $\eps$ (so that the original 
solution is recovered at $\eps=1$) which ``magnifies'' a vicinity of 
this point. In the limit when $\eps\rightarrow 0$ we get the 
``near horizon limit'' which is described by the metric  
\bea
ds^2&=&-2{\hat H}^{-1}(du+\beta_mdx^m)\left(dv+{\hat\omega}_m dx^m+
\frac{F}{2}(du+\beta_mdx^m)\right)+{\hat H}h_{mn} dx^mdx^m
\eea
with ${\hat\omega}$ given by a solution of (\ref{FlatToAdS}). To see how 
the replacement $\omega\rightarrow{\hat\omega}$ arises, we look at
\bea\label{OmToHatOM1}
dv'+\omega'=dv'+{\hat\omega}'+(\omega'-{\hat\omega}')
\eea
We now recall that by definition (\ref{FlatToAdS}), the form 
$\omega'-{\hat\omega}'$ satisfies the equation
\bea
[d(\omega'-{\hat\omega}')]^+=
\frac{1}{H'}(d\omega')^++\frac{1}{2}\frac{F'}{H'}d\beta'\sim
\frac{1}{H}(d\omega)^++\frac{1}{2}\frac{F}{H}d\beta
\eea 
where we used $\sim$ to denote the leading order at small $\eps$.
Introducing ${\tilde\omega}$ as a solution of the equation
\bea
(d{\tilde\omega})^+=\frac{1}{H}(d\omega)^++\frac{1}{2}\frac{F}{H}d\beta
\eea 
we can rewrite (\ref{OmToHatOM1}) as
\bea
dv'+\omega'=\frac{1}{\eps^2}
\left[dv+{\hat\omega}+\eps^2{\tilde\omega}+\dots\right]
\eea
So indeed in the limit $\eps\rightarrow 0$, $\omega$ is replaced by 
$\hat\omega$.

To summarize, we have shown that in the three charged system there 
exists a limit (\ref{NearHorLimit}) which reproduces the ``near horizon 
map'' (\ref{FlatToAdS}). However, we want to stress again that this 
limit only magnifies a vicinity of a particular point where $H$ 
has singularity, while solution with ${\hat H}$ and ${\hat\omega}$ 
can be extended beyond such vicinity. In this sense, one can view the map 
(\ref{FlatToAdS}) as a solution generating technique which was inspired 
by the near horizon limit.

\section{Elimination of coordinate singularities.}
\label{AppSingul}

In this appendix we will provide some extra details on how the apparent 
singularities of the solution (\ref{NewHhat})--(\ref{NewBase}), 
(\ref{MySlnMetric}), (\ref{MySlnFlat})  can be eliminated by an 
appropriate change of coordinates. While the basic ideas of such 
reparameterizations were outlined in section \ref{SectProp}, there we 
skipped 
some technical details which will be explained in this appendix.

This appendix consists of two main parts: one deals with singularities at 
$r=0$, but $f\ne 0$, and another deals with vicinity of regions where 
$f=0$ without any limitation on the value of $r$. This is a very natural 
split, since $r=0$ was a singularity associated with our 
choice of spherical coordinates, while $f=0$ seemed to be a real 
singularity where the harmonic functions had their sources. We will 
show that contrary to a naive expectation, the solution is 
completely regular at those points as well.

{\bf 1. Singularity at $r=0$.}

We already did a partial analysis of this singularity in section 
\ref{SectProp}, here we will present some details we skipped before, 
and we also analyze the poles on the sphere ($\theta=0$ of 
$\theta=\pi/2$) when they located at $r=0$.     

We begin with generic case ($\theta\ne 0$, $\theta\ne \pi/2$). As we 
already mentioned in section \ref{SectProp}, the metric of the base space 
(\ref{ApproxBase}) becomes regular after the change of variables\footnote{
We will also need the inverse relation between angles: 
$\phi={\tilde\phi}+\nu{\tilde\psi}$, 
$\psi={\tilde\phi}+(\nu+1){\tilde\psi}$.}
(\ref{RegREq0})
\bea
{\tilde\phi}=(\nu+1)\phi-\nu\psi,\quad {\tilde\psi}=\psi-\phi,\qquad
x_1=r\cos{\tilde\psi},\quad x_2=r\sin{\tilde\psi}.
\eea
Here we will show that this change of variables does not introduce 
any new singularities into $g_{\mu\nu}$. We already know that there 
are no singularities in $h_{mn}$, so we have to analyze
\bea
du+\beta\approx du+\frac{Q}{a\sqrt{2}}\left[-d{\tilde\psi}+
\frac{\cos 2\theta}{(\nu+1)\cos^2\theta-\nu\sin^2\theta}
d{\tilde\phi}\right]
\eea
and
\bea
dv+\omega&\approx& dv+\frac{Q}{a\sqrt{2}}\left[d{\tilde\psi}+
\frac{(\nu+\cos^2\theta)d{\tilde\phi}}{[(\nu+1)\cos^2\theta-
\nu\sin^2\theta]^2}
\right]\nonumber\\
&+&
\sqrt{2}a\nu(\nu+1)\left[-d{\tilde\psi}+
\frac{\cos 2\theta}{(\nu+1)\cos^2\theta-
\nu\sin^2\theta}d{\tilde\phi}\right]
\eea
To have regular metric we have to avoid an appearance of $d{\tilde\psi}$ 
unless it is multiplied by $r$. Clearly this can be achieved by 
diffeomorphism (\ref{R0Twist}). This concludes our analysis of the 
singularity at $r=0$ and generic values of $\theta$, we now proceed with 
analysis of the poles on the sphere.

First we look at $r=0$, $\theta=0$. Near this point the metric 
(\ref{ApproxBase}) simplifies:
\bea
h_{mn}dx^m dx^n\approx (\nu+1)(dr^2+a^2 d\theta^2)+
\frac{a^2\theta^2}{\nu+1}
\left[(\nu+1)d\phi-\nu d\psi\right]^2+\frac{r^2}{\nu+1}d\psi^2
\eea
For the relevant one--forms we get:
\bea\label{BadPrfUV}
du+\beta&\approx& du+\frac{Q}{a\sqrt{2}}\left[-\frac{d\psi}{\nu+1}+
\frac{\theta^2d\phi}{\nu+1}\right],\nonumber\\
dv+\omega&\approx& dv+\frac{Q}{a\sqrt{2}}
\left[\frac{1+4\nu}{\nu+1}\theta^2 d\phi+\frac{d\psi}{\nu+1}\right]
+a\sqrt{2}\nu\left[-d\psi+\theta^2d\phi\right]
\eea
and the scalar functions are
\bea
H=1+\frac{Q}{a^2(\nu+1)},\qquad F=-2\nu.
\eea
Substituting this in (\ref{MySlnMetric}) we get an approximate metric 
near the points
where $r=0,\theta=0$:
\bea\label{MetrNearCone}
ds^2&=&-\frac{[Q+(\nu+1)a^2][Q-\nu(\nu+1)a^2]}{(\nu+1)a^2} d{\tilde U}d{\tilde V}
\nonumber\\
&+&\frac{(\nu+1)Q}{Q+(\nu+1)a^2}(dr^2+a^2d\theta^2)
+\frac{(\nu+1)a^2Qr^2}{[Q+(\nu+1)a^2][Q-\nu(\nu+1)a^2]^2}dy^2\nonumber\\
&+&\frac{(\nu+1)Qa^2\theta^2}{Q+(\nu+1)a^2}\left\{d\phi+
\frac{\nu(\nu+1)a^3}{Q[Q-\nu(\nu+1)a^2]}dy\right\}^2\\
&+&r^2 dy^2(A_1 d{\tilde U}+A_2d{\tilde V})+
\theta^2\left\{d\phi+
\frac{\nu(\nu+1)a^3}{Q[Q-\nu(\nu+1)a^2]}dy\right\}
(A_3 d{\tilde U}+A_4d{\tilde V})\nonumber
\eea
Here we introduced ${\tilde U}$ and ${\tilde V}$ as some 
special linear combinations of $t,y,\phi,\psi$ with constant 
coefficients. Neither explicit form these combinations nor 
the expressions for the constants $A_1$, $A_2$, $A_3$, $A_4$ 
are important for our discussion, so we do not write them 
down to avoid unnecessary complications.

The expression (\ref{MetrNearCone}) clearly demonstrates that 
all curvature invariants stay finite as we {\it approach} the 
point $r=0,\theta=0$. However we still can encounter a conical 
singularity located exactly at that point. To see this let us 
look at the following two terms:
\bea
&&\frac{(\nu+1)Q}{Q+(\nu+1)a^2}dr^2
+\frac{(\nu+1)a^2Qr^2}{[Q+(\nu+1)a^2][Q-\nu(\nu+1)a^2]^2}dy^2\nonumber\\
&&\qquad=
\frac{(\nu+1)Q}{Q+(\nu+1)a^2}\left\{dr^2+r^2
\left(\frac{ady}{Q-\nu(\nu+1)a^2}\right)^2
\right\}
\eea
This metric has a conical singularity unless we make 
an identification $y\sim y+2\pi R$ with 
\bea\label{CorrectRApp}
R=\frac{Q-\nu(\nu+1)a^2}{a}
\eea
Suppose we make such identification (we recall that coordinate $y$ 
was periodic to begin with and this identification simply chooses 
a particular relation between the values of $R$ and $a$). Then we 
can trade a parameter $a$ in (\ref{MetrNearCone}) for the radius of
$y$ circle\footnote{We still kept $a$ in some 
constants in the metric (\ref{MetrNearCone1}). We implicitly 
assume that $a$ entering those constants is expressed in terms of $R$
by solving a quadratic equation coming from (\ref{CorrectRApp}).}:
\bea\label{MetrNearCone1}
ds^2&=&-R\frac{Q+(\nu+1)a^2}{(\nu+1)a} d{\tilde U}d{\tilde V}
+\frac{(\nu+1)Q}{Q+(\nu+1)a^2}\left(dr^2+r^2\frac{dy^2}{R^2}\right)
\nonumber\\
&+&\frac{(\nu+1)Qa^2}{Q+(\nu+1)a^2}\left(
d\theta^2+\theta^2\left\{d\phi+
\frac{\nu(\nu+1)a^2}{Q}\frac{dy}{R}\right\}^2\right)\\
&+&r^2 dy^2(A_1 d{\tilde U}+A_2d{\tilde V})+
\theta^2\left\{d\phi+\frac{\nu(\nu+1)a^2}{Q}\frac{dy}{R}\right\}
(A_3 d{\tilde U}+A_4d{\tilde V})\nonumber
\eea
We see that in order to avoid a conical singularity at $\theta=0$, 
the relation 
\bea
\frac{\nu(\nu+1)a^2}{Q}=m
\eea
should be satisfied for some integer $m$. 

To summarize, we found that our solution has regular curvature 
invariants as we approach a point $r=0,\theta=0$, but at that point 
itself the solution generically has a conical singularity. However if 
the parameters of the solution satisfy two relations
\bea\label{RegCondApp}
R=\frac{Q-\nu(\nu+1)a^2}{a},\qquad a^2=\frac{Qm}{\nu(\nu+1)}
\eea
for some integer $m$, then the solution is completely regular. 

Now we look at the vicinity of a point where 
$r=0,\theta=\frac{\pi}{2}$. Fortunately, we don't have to do 
any new analysis there since the solution has a $Z_2$ symmetry:
\bea
\theta\rightarrow \frac{\pi}{2}-\theta,\quad
\phi\rightarrow -\psi,\quad \psi\rightarrow -\phi,\quad
\nu\rightarrow -(\nu+1)
\eea
This symmetry shows that our solution is regular at 
$r=0,\theta=\frac{\pi}{2}$ as long as (\ref{RegCondApp}) is satisfied.

{\bf 2. Singularity at the sources of harmonic functions.}

Let us now analyze the vicinity of points where the coefficient 
functions $(H,F,\beta,\omega,h_{mn})$ diverge. This happens in 
the points where $f=0$. We will use the following trick. Instead of 
$H$ and $\omega$ we introduce the following functions:
\bea
H_{\alpha}=\alpha+\frac{Q}{f},\qquad 
\omega_\alpha={\hat\omega}+\alpha(\omega-{\hat\omega})=
{\hat\omega}+\alpha\frac{2\nu(\nu+1)a^2}{Q}\beta
\eea
Obviously $\alpha=1$ corresponds to the original functions, 
while $\alpha=0$ corresponds to the near horizon functions
${\hat H}$ 
and ${\hat\omega}$. We also notice that the terms proportional to 
$\alpha$ in $H_{\alpha}$ and $\omega_\alpha$ become less and less 
relevant as one approaches the singularity (i.e. as $f\rightarrow 0$), 
so one would hope that very close to singularity the metric is almost 
the same as for the near horizon case, and the contributions with 
various powers of $\alpha$ would be suppressed. We will show that the 
terms containing higher powers of $\alpha$ can indeed be treated as small 
perturbation of the metric with $\alpha=0$. Since that metric was 
regular by construction (see (\ref{AfterSF})), the perturbation does not 
destroy regularity, and this will serve as a proof of the desired result. 

Let us now implement this strategy. We write the metric corresponding to 
the solution $(H_\alpha,F,\beta,\omega_\alpha,h_{mn})$ keeping only the 
terms $\alpha^0$ and $\alpha^1$:
\bea
ds_\alpha^2&=&-2H_\alpha ^{-1}(du+\beta)\left(dv+\omega_\alpha+
\frac{F}{2}(du+\beta)\right)+H_\alpha h_{mn} dx^mdx^m\nonumber\\
&=&ds_0^2+2\alpha\frac{f^2}{Q^2}(du+\beta)\left(dv+{\hat\omega}+
\frac{F}{2}(du+\beta)\right)+\alpha h_{mn} dx^mdx^m\nonumber\\
&&-4\alpha \nu(\nu+1)\frac{a^2f}{Q^2}\beta(du+\beta)\\
&=&\left(1+\frac{\alpha f}{Q}\right)ds_0^2+
4\alpha\frac{f^2}{Q^2}(du+\beta)\left(dv+{\hat\omega}+
\frac{F}{2}(du+\beta)-\nu(\nu+1)\beta\frac{a^2}{f}\right)\nonumber
\eea
In this expression $ds_0$ denotes the metric (\ref{AfterSF}) which was 
completely regular everywhere including the region which we are 
considering. The bracket in the first term describes a perturbation
which goes to zero as we approach the ``singularity'' and thus it does 
not spoil the solution. In the second term the contributions containing 
$du$ or $dv$ go to finite limits as $f$ goes to zero, however there 
is a potentially dangerous piece proportional to $1/f$:
\bea
4\alpha\frac{f^2}{Q^2}\beta\left({\hat\omega}+
\frac{F}{2}\beta-\nu(\nu+1)\beta\frac{a^2}{f}\right)
\eea 
Miraculously the singular contributions in this term cancel out. To see 
this we rewrite ${\hat\omega}$ in the form:
\bea
{\hat\omega}=\frac{2a^2}{f}\nu(\nu+1)\beta+
\frac{aQ}{\sqrt{2}f}(2\nu+1)(\cos^2\theta d\psi+\sin^2\theta d\phi)
\eea
Then we find:
\bea
4\alpha\frac{f^2}{Q^2}\beta\left({\hat\omega}+
\frac{F}{2}\beta-\nu(\nu+1)\beta\frac{a^2}{f}\right)=
4\alpha\frac{f^2}{Q^2}\beta\left\{\frac{aQ}{\sqrt{2}f}(2\nu+1)
(\cos^2\theta d\psi+\sin^2\theta d\phi)\right\}
\eea
and we see that the singular piece indeed disappears and we have a 
regular perturbation of the metric near a point where $f=0$: 
\bea
ds_\alpha^2
&=&\left(1+\frac{\alpha f}{Q}\right)ds_0^2+
\frac{4\alpha}{Q^2}(fdu+f\beta)\left(fdv+
\frac{aQ}{\sqrt{2}}(2\nu+1)
(\cos^2\theta d\psi+\sin^2\theta d\phi)\right)\nonumber
\eea

We now look at the subleading orders in $\alpha$. The term proportional 
to $\alpha^2$:
\bea
&&-2\frac{\alpha^2 f^3}{Q^3}
(du+\beta)\left(dv+{\hat\omega}+
\frac{F}{2}(du+\beta)\right)+
4\nu(\nu+1)\frac{\alpha^2 a^2 f^2}{Q^3}(du+\beta)\beta\\
&&\qquad=-2\frac{\alpha^2 f^3}{Q^3}(du+\beta)\left(
dv+\frac{aQ}{\sqrt{2}}(2\nu+1)
(\cos^2\theta d\psi+\sin^2\theta d\phi)-\nu(\nu+1)\frac{a^2}{f}\beta
\right)\nonumber
\eea
goes to finite limit as $f\rightarrow 0$, and all other terms vanish 
as we approach the ``singularity''. Indeed, the contribution of the order 
$\alpha^k$ looks like
\bea
(-1)^k\left[-2\frac{\alpha^k f^{k+1}}{Q^{k+1}}
(du+\beta)\left(dv+{\hat\omega}+
\frac{F}{2}(du+\beta)\right)+
4\nu(\nu+1)\frac{\alpha^k f^k}{Q^{k+1}}(du+\beta)\beta\right]
\eea
so it goes to zero as $f^{k-2}$ for $k\ge 3$.

Let us now collect the terms with all powers of $\alpha$ and resum the 
series to produce an {\it exact} metric:
\bea\label{SumPertMetr}
ds^2_\alpha=\left(1+\frac{\alpha f}{Q}\right)ds_0^2+
\frac{4\alpha}{Q^2}(fdu+f\beta)\left(fdv+
\frac{aQ}{\sqrt{2}}(2\nu+1)
(\cos^2\theta d\psi+\sin^2\theta d\phi)\right)
\nonumber\\
-2\frac{\alpha^2 f^3}{Q^3}\left(1+\frac{\alpha f}{Q}\right)^{-1}
(du+\beta)\left(
dv+\frac{aQ}{\sqrt{2}}(2\nu+1)
(\cos^2\theta d\psi+\sin^2\theta d\phi)-\nu(\nu+1)\frac{a^2}{f}\beta
\right)
\eea
The series which we had to compute is a simple geometric series and 
it converges when $f$ becomes small enough (i.e. if 
$\frac{\alpha f}{Q}<1$).
We can now take $\alpha=1$ (and we can always go close enough to the 
surface where $f=0$, so that the geometric series converges), then from 
the analysis 
presented above, we can see that the metric is completely regular apart 
from possible conical defects. The location of conical defects is 
dictated by the structure of $ds_0^2$ and (if $\nu$ is not an integer)
these defects could indeed appear at either $(r=0,\theta=0)$ or 
$(r=0,\theta=\frac{\pi}{2})$. Fortunately for $\nu\ne 0$ and $\nu\ne -1$ 
the function $f$ does not vanish at those points, and thus according to 
the 
analysis in the first part of this appendix, even conical defects do
not arise. 

It is very important that 
we always used the original coordinates, so the regularity of the metric 
$g_{\mu\nu}$ near $f=0$ also implies the regularity of the inverse 
metric and thus the regularity of all curvature invariants. To see this 
we recall that the determinant of the metric (\ref{detmet}) goes to 
finite limit as we approach $f=0$, so the inverse metric is indeed 
regular. In particular, $g^{tt}$ does not blow up as we approach $f=0$, 
so this is not a surface of infinite redshift.

To summarize, we have shown that near any point where $f=0$, the metric 
can be decomposed into a regular $AdS_3\times S^3$ geometry $ds^2_0$ 
which comes from the near horizon limit, and smooth corrections so 
that the entire metric is completely regular.

\end{document}